\documentclass[12pt]{article}
\usepackage{amsmath}
\usepackage{amssymb}
\usepackage{epsfig}
\usepackage{cite}
\usepackage{color,colordvi}

\begin{document}
\vspace{0.01cm}
\begin{center}
{\Large\bf  Black Hole Macro-Quantumness } 

\end{center}

\vspace{0.1cm}

\begin{center}

{\bf Gia Dvali}$^{a,b,c}$ and {\bf Cesar Gomez}$^{a,e}$\footnote{cesar.gomez@uam.es}

\vspace{.6truecm}


{\em $^a$Arnold Sommerfeld Center for Theoretical Physics\\
Department f\"ur Physik, Ludwig-Maximilians-Universit\"at M\"unchen\\
Theresienstr.~37, 80333 M\"unchen, Germany}


{\em $^b$Max-Planck-Institut f\"ur Physik\\
F\"ohringer Ring 6, 80805 M\"unchen, Germany}

%


{\em $^c$Center for Cosmology and Particle Physics\\
Department of Physics, New York University\\
4 Washington Place, New York, NY 10003, USA}

{\em $^e$
Instituto de F\'{\i}sica Te\'orica UAM-CSIC, C-XVI \\
Universidad Aut\'onoma de Madrid,
Cantoblanco, 28049 Madrid, Spain}\\

\end{center}


\begin{abstract}
\noindent  
 
{\small

It is a common  wisdom  that  properties of macroscopic bodies are well described by (semi)classical physics.   As we 
have suggested \cite{Nportrait, Hair, Quantum} this wisdom is not applicable to black holes.
 Despite being macroscopic,  black holes are quantum objects.  They represent Bose-Einstein condensates of $N$-soft gravitons 
 at the quantum critical point, where $N$ Bogoliubov modes become gapless. 
  As a result,  physics governing arbitrarily-large black holes (e.g., of galactic size) is a quantum physics of the collective Bogoiliubov modes. 
  This fact introduces a new intrinsically-quantum corrections in form of $1/N$, as opposed to 
 $e^{-N}$. These corrections  are unaccounted by the usual  semiclassical expansion in $\hbar$ and  cannot 
 be recast in form of a quantum  back-reaction to classical metric.  Instead the metric itself becomes  an approximate 
 entity. These $1/N$ corrections abolish the presumed properties of black holes, such as non existence of hair, and 
 are the key to nullifying the so-called  information paradox. }

%
%

\end{abstract}

\thispagestyle{empty}
\clearpage

\section{Essence of Macro-Quantumness}

 It is a common wisdom that the properties derived in idealized semi-classical treatment, such as, e.g., Hawking's exact thermality \cite{hawking} \cite{bekenstein} and absence of hair \cite{no-hair},  must be  well-applicable to the real macroscopic black holes. 
 From the first glance, this sounds reasonable.  After all, the common 
 effective-field-theoretic sense tells us that for large objects all the microscopic quantum physics averages out in effective macroscopic characteristics, which 
 are classical.  When applying this reasoning to ordinary macroscopic objects 
 such as planets, stars or galaxies, no apparent  paradoxes or inconsistencies 
 appear.  For example, treating the earth as a semi-classical gravitating source  
gives a consistent picture. 

 In contrast, when applying the same common sense to realistic macroscopic black holes of finite mass, one ends up with puzzles and paradoxes, perhaps the most prominent being 
 Hawking's information paradox \cite{Haw}.   
 The purpose of this short note is not to discuss  the existing puzzles one by one, but instead to point out the misconception that underlies all of them. 
 Namely, the quantum effects for the macroscopic black holes are much more important than what is suggested by straightforward application of semi-classical reasoning.   This is the lesson from the recently-developed black hole quantum portrait \cite{Nportrait, Hair, Quantum, cosmo}.  In this respect there is nothing new in the present note, but 
 we shall provide a sharper focus and specifically reiterate the key point  that we believe sources the black hole mysteries.   
  We would like to explain why it happens that some macroscopic bodies are more quantum than others.   

 The short answer is that despite being macroscopic, black holes are systems at 
 the critical point of a {\it quantum} phase transition \cite{Quantum}.  As a result, no matter how large and heavy, they can never be treated fully classically. Indeed the very nature of the phase transition as quantum requires a great amount of quantumness in the form of entanglement.
Of course, for some aspects (e.g., large distance gravitational effects on probe bodies) the semi-classical treatment is fine, but is non-applicable for other aspects, such as information storage and processing.  

 In order to explain this profound difference, let us compare a black hole to an ordinary macroscopic object, e.g., a planet or a bucket of water.  Of course the common property of all the macroscopic objects, that allows to treat them in long-distance regimes classically, is the large number of quantum constituents, $N$.   
  Property, $N \, \gg \, 1$ is universally shared by all the macroscopic bodies of our interest. 

 For such objects we can define some quantum characteristics, such as, $N$ (e.g., number of atoms 
 in the bucket of water) and their {\it quantum} coupling strengths.  However, in ordinary objects with size much bigger than the de-Broglie wave-lengths of the constituents,  the coupling $\alpha_{ij}$ 
 between a pair of constituents $i$ and $j$ strongly depends on the relative positions of the constituents (for example, a nearest neighbor coupling in atomic systems ) and cannot be defined 
 universally.   
 
 In contrast, an universal coupling can be defined in the systems in which everyone talks to everyone 
 at an equal strength. Such is the property of Bose-Einstein condensates (BECs) where all the constituents are in a common quantum state.  In particular, 
 the black holes represent such condensates of gravitons.   For such a system we can define a very 
 useful parameter,  
 \begin{equation}
 \lambda \equiv N \alpha \, , 
 \label{thooft}
\end{equation} 
 which is somewhat analogous to the 't Hooft coupling for 
 gauge theories with $N$-colors \cite{t'Hooft}.  Despite the crucial difference that in our case 
 $N$ is not an input of the theory, but rather a characteristic of a particular 
 BEC, we shall refer to $\lambda$ as the 't Hooft coupling. 
 
    This parameter plays the central role in our considerations 
 since it determines how close is the system from {\it quantum} criticality.  Thus,  level of classicality of the system is not determined by only how large $N$ is, but most importantly how far it is from the  critical value $N \, = \, 1/\alpha$.  This is the fundamental difference between black holes  
and other macroscopic bodies with many constituents.  For the ordinary macroscopic objects, such as 
planets,  the analog of the 't Hooft's coupling either cannot be defined or it is far from quantum criticality. 
This is why the ordinary macroscopic objects can be treated classically with a very good approximation, 
without encountering any seeming paradoxes.  Contrary, as we have shown,  black holes are always 
at the quantum critical point $N\alpha \, = \, 1$ up to $1/N$ corrections. As a result, black holes can never be treated classically.  There are certain quantum effects (such as mass gap and degeneracy of Bogoliubov modes) that for large black holes become extremely important.  In particular, at the quantum critical point small subsystems are maximally entangled i.e the entanglement entropy for the reduced one particle density matrix is maximal.

\newpage

 We thus, have outlined the following sequence of macroscopic systems with 
 increasing level of quantumness:
 
 $~~~~~$
 
 \begin{center}
 {\bf  Ordinary macroscopic objects (e.g., planets or buckets of water).} 
  
    Quantum Characteristics : $N$ exists,  $\lambda$ cannot be defined.  
  
  $\downarrow$

 {\bf  Generic (non-critical) Bose-Einstein-Condensates. }
  
  Quantum Characteristics : Both $N$ and $\lambda$  are well defined, 
  but $\lambda \neq 1$.   
  
   $\downarrow$
  
 {\bf  Black holes: Bose-Einstein condensates stuck at the quantum critical point.}  
 
 Quantum Characteristics : Both $N$ and $\lambda$  are well defined, 
  and  $\lambda =1$.   
 
\end{center} 
 
 $~~~~$

  In order to explain this profound difference, let us consider a hypothetical gravitating source 
of the mass of a neutron star. We shall use an oversimplified model in which we shall approximate the 
source by a collection of $N_B$  particles  of baryonic mass, $m_B \sim$  GeV, stabilized by some non-gravitational forces. We shall ignore the contribution to the energy from the stabilizing force. 
 Then by dialing the strength of the stabilizing force, we can bring the system to the critical point of 
 black hole formation.  In the classical approximation such a "neutron star" outside produces 
 a gravitational field identical to the one produced by a classical Schwazschild black hole. 
 So why is the case that for the neutron star the quantum effects are not important 
 whereas for a black hole of the same mass they are absolutely crucial? 
 
     In order to answer this  question let us reduce the quantum portrait of the above system to its bare essentials.  We are dealing with a source, represented by a multi-baryon state of occupation number 
 $N_B \, \sim \, 10^{57}$ and size $L \, \sim \, 10^6$ cm. This source is not a Bose-Einstein condensate, since baryons (even if spin-$0$) are not  in the same state, and in particular their de Broglie wave-lengths are much shorter than the size of the system.  However, these baryons source gravity and produce gravitational field that 
 contains approximately  $N \, \sim \, 10^{77}$ gravitons.  
  The two occupation numbers are related as, 
 \begin{equation}
      N \, = \, N_B^2  (m_B / M_P)^2 \, , 
  \label{numbers}
 \end{equation}
 where  $M_P$ is the Planck mass, and we shall also define the Planck
length $L_P \, \equiv \, \hbar /M_P$.
 Unlike baryons, these gravitons are much closed to being 
 a Bose-Einstein condensate, because  the majority of them occupy the same state, and in particular have comparable characteristic wave-lengths $L$ given by the size of the baryonic source, $L \, \sim \,  L_{star}$. 
 Due to this, in contrast to the baryonic constituents of the star, for gravitons we can define an universal quantum 
 coupling, 
  \begin{equation}
      \alpha \, \equiv \, (L_P / L)^2 
  \label{alpha}
 \end{equation}
 and the corresponding 't Hooft's coupling $\lambda$ given by 
 (\ref{thooft}). 
   The only caveat is that the graviton condensate is not self-sustained
as long as $L_{star} \, > \, r_g$.   
 That is, the gravitational mass (self-energy) of the graviton condensate  $M_{gr} \, = \, N \hbar /L$ is below the mass of the baryonic source 
   $M_{star} \, = \, N_B m_B$ and alone is not enough to keep the gravitons together.  Classically, we think of this situation as 
   the size of the source $L_{star}$ being larger than the corresponding gravitational radius 
   $r_g \, \equiv  \, M_{star} L_P^2/\hbar$, but we see that the quantum-mechanical reason is that 
   the 't Hooft coupling is far from criticality. Indeed, expressing $N$ and $\alpha$ through their 
  dependence on $L$ and $r_g$, we have, 
   \begin{equation}
     \lambda \, \equiv \,  N \alpha \, =  \, (r_g / L)^2 \, = \, (r_g / L_{star})^2 \, .
  \label{Nalpha}
 \end{equation}
 
  Thus, the classical statement that a given source is not a black hole ($r_g \, < \, L_{star}$), 
  quantum-mechanically translates as the condition  that  the 't Hooft coupling 
  of graviton condensate is weak, $\lambda < \, 1$.    Thus, the standard 
  semi-classical expansion in powers of $r_g/L$ is nothing but an expansion in the 't Hooft 
  coupling $\lambda$.  This expansion ignores additional $1/N$-effects. That is,  
 it represents a planar approximation:
 \begin{equation}
  \lambda \, = \, {\rm fixed}, ~~~  N = \infty \, . 
 \label{planar}
 \end{equation}
   Such approximation is justified 
  {\it only}  as long as $\lambda$ is sub-critical.    
 
 Now imagine that by changing the parameters of the model (say, by decreasing a stabilizing force) we 
 bring the source to  the point $L_{star} \,  = \,  r_g$. Classically, we think of this point as a point 
 of classical black hole formation, but in reality this is a critical point of a {\it quantum} phase transition!
  As we have shown \cite{Quantum},  there are dramatic quantum effects which take place at this point. 
 In particular,  of order $N$ Bogoliubov modes of the graviton condensate become gapless and nearly degenerate.  The condensate starts 
a quantum depletion, leakage and a subsequent collapse.  This is the underlying  
quantum-mechanical nature of the process that semi-classically is viewed as Hawking evaporation.  But,   Hawking's semi-classical limit in our language corresponds to planar limit, in which only $\lambda$-corrections are kept 
whereas  $1/N$-corrections are not taken into the account.   
   In reality every act of emission differs from this idealized  approximation by $1/N$-corrections.  Our point is to stress the extreme importance of these corrections.

    In other words for a generic BEC the quantity $1/N$ measures the {\it quantum noise} of the system. For $N \gg 1$ these effects can be thought as very tiny and effectively negligible. This is in fact the case provided the constituents of the system are not entangled. However, and this is the key of the quantum phase transition, quantum noise makes a dramatic difference when the constituents are maximally entangled i.e at the quantum critical point. In fact at this point  the entanglement entropy for the reduced one particle density matrix becomes maximal and a new branch of light Bogoliubov modes appear \cite{students}. 
     This is something completely alien to any classical system.  In this sense black holes are intrinsically quantum objects.   This phenomenon is fully missed in classical or semi-classical analysis. Its discovery requires a microscopic quantum view.  
  
   {\it  Thus, even macroscopic black holes are quantum.} 
   
    This is a very general message we wanted to bring across in this short note.

\section{Quantumness Versus Semi-Classicality}

 Can the quantum effects we are pointing out be  somehow read off in 
 the standard semi-classical treatment?   We shall now explain why the answer is negative.  
 
  In standard treatment the  black holes  are introduced through the metric
  $g_{\mu\nu}(x)$, which is an intrinsically-classical entity. 
  The effects of quantum gravity are then thought to be accounted  in terms of 
 quantum corrections to metric, without abandoning the very concept of the (classical) metric.  In other words, both before and after the quantum corrections 
 the metric itself is treated as a background classical field. The role 
 of the quantum gravity  is reduced to understanding  the rules of corrections according to which this classical entity changes, without  abolishing the very concept  
 of a background metric.   We claim that for certain macroscopic systems,  such  
 as  black holes,  the above treatment is inconsistent.   
     
 It is absolutely crucial to understand that $1/N$-corrections are intrinsically-quantum and can never be recast in form of some quantum-back-reacted metric.
 Instead the very notion of the metric needs to be abandoned and be treated as  
 approximate.  
 In order to explain  this, let us go through the three levels of quantumness: 
 
 \begin{center}
  {\bf  Classical:  $\hbar =  0, ~ {1 \over N } =  0$;
 
 $\downarrow$
   
  Semi-Classical:  $\hbar \, \neq \, 0, ~ \, {1 \over N } =  0$ ; 

 $\downarrow$

   Quantum:  $\hbar \, \neq \, 0, ~ {1 \over N } \neq 0$ . }
 \end{center}

 Consider a light test body and a heavy source of energy momentum tensors 
$\tau_{\mu\nu}$ and $T_{\mu\nu}$ respectively.  
 In classical  GR a scattering of a probe on a source can be understood in terms of a propagation of the former in a background classical metric created by the  latter,  with an amplitude, 
 \begin{equation}
      A_{Cl} \, = \, \int_x  \, g_{\mu\nu}(x)\tau^{\mu\nu}(x)\, , 
\label{amplitude}
\end{equation}
where integration is performed over a four-dimensional space-time volume. 
 The metric $g_{\mu\nu}$ is obtained by solving the classical Einstein equation 
 with the source $T_{\mu\nu}$.  It is well-known that exactly the same 
 amplitude can be reproduced by summing up the infinite series of tree-level Feynman  diagrams with intermediate graviton lines, 
 
 \begin{eqnarray}
  A \, &= &\,  G_{N} \int_{x,y} \, T(x) \Delta(x-y) \tau(y) \, + \, \nonumber \\ 
     &&G_N^2  \int_{x,y,z,w} T(x) T(y) \Delta(x-w)
     \Delta(y-w) \Delta(z-w) O(w)  \tau(z) \, + \, ...  \nonumber \\
\end{eqnarray} 
 Here $\Delta(x)$ is a graviton propagator, and tensorial indexes are suppressed.  These  series are non-zero despite 
 the fact that we are working in $\hbar = 0$ limit, and they fully reproduce 
 the result obtained by considering the motion in the classical metric (\ref{amplitude}).  In fact,  order by order the above series reproduce the expansion of  a classical solution of Einstein equation in series of 
 $G_N$.  For example, for  a spherical  source of mass $M$, the above series 
 reproduce the expansion of Schwarzschild metric in series of  $r_g \over r$ where $r_g\equiv 2G_NM$ is the gravitational radius of the source and $r$ is a radial coordinate \cite{Duff}.  
 
  Let us now move towards the quantum picture, $\hbar \neq 0$.  
  The standard idea about how to take into the account 
 quantum gravity effects is to integrate out loops and write down the $\hbar$-corrected 
 effective action for $g_{\mu\nu}(x)$.  The action obtained in this way will 
 in general contain an infinite series of curvature invariants, with each power of 
 curvature being accompanied by a factor of $L_P^2$ (in absence of other input scales).  The effective 
 quantum-corrected metric is then represented as a solution to the equations 
 obtained by varying the effective action. 
  In this philosophy, the quantum gravity effects are accounted in form of a back reaction to the classical metric.  The quantum-corrected metric obtained in such a way, although formally includes $\hbar$-effects is still treated as a classical entity:
  \begin{equation}
 g_{\mu\nu}(x) \, \rightarrow \,  g_{\mu\nu}' \, = \,  g_{\mu\nu}(x) \, + \, \delta g_{\mu\nu}(x, \hbar)  \, .
  \label{newmetric}
  \end{equation}
 In particular, the quantum-corrected  scattering of a probe over the source in this limit 
 can still be reduced to the effects of propagation in the background 
 metric obtained  by replacing in (\ref{amplitude})  $g_{\mu\nu}$ by $g_{\mu\nu}'$.   
  
  As a result, such an analysis is not really quantum, but rather {\it semi-classical}, as 
  it never resolves the quantum constituents of the metric ($1/N \, = \, 0$).

{\it  This is the essence of semi-classical approximation:  It reduces quantum effects 
 to the $\hbar$-correction of classical entities, without resolving their constituency. } 
 
  We thus claim, that the above treatment of quantum gravity misses out 
 the  $1/N$-corrections, which are absolutely crucial for black holes.
  The physics generated by these corrections, is impossible to be reproduced by any quantum corrections to the classical metric.  Instead,  the very notion of the metric must be abandoned and only treated as approximate.

 In our language it is clear why this is the only consistent treatment. 
Indeed, it is impossible to keep all three quantities $M, L_P$ and $\hbar$ 
finite, and simultaneously keep $1/N=0$.   
  Putting it differently,   $r_g$-corrections are corrections in terms of 
 series in 't Hooft coupling  $\lambda=(\alpha N)$, which are 
 different from $1/N$-series.  Naively, it seems that one can consistently 
 keep the former while discarding the latter 
 by taking the planar limit (\ref{planar}).  However, this is an illusion,  since
 in this limit also  the black hole evaporation time ( which scales as $N^{3/2}$) becomes infinite,  so that the integrated effect is still finite.   

  Notice, that $1/N$ corrections are present already in the tree-level 
 scattering of a probe over a black hole and come from the processes 
 in which the probe exchanges the momentum with individual constituent   
 of the graviton condensate.  Because the condensate is at the quantum critical 
 point, such exchanges cost $1/N$ as opposed to $e^{-N}$. 
 
   The resulting quantum scattering amplitude $A_Q$ differs from its  
   classical counterpart by $1/N$-effects, 
  \begin{equation}
   A_Q \, = \, A_{Cl} \, + \,  {\mathcal O} (1/N) \, .   
  \label{correct}
  \end{equation}
  However, the crucial point is that, unlike the semi-classical case, these effects  cannot be recast in form of propagation in any new  corrected metric. That is, the 
  quantum amplitude  $A_{Q}$
  does not admit any representation in form of 
  \begin{equation}
   A_Q \, = \, \int_x  g_{\mu\nu}(x)' \tau^{\mu\nu} \, , 
 \label{correct}
 \end{equation} 
 where $g_{\mu\nu}(x)'$ could be any sensible metric.   Such representation of the amplitude ceases to exist as soon as we correctly account for $1/N$-effects.
 \footnote{The quantity $N$ defined in \cite{Nportrait} as a measure of classicality also emerges in \cite{ramy}. However,  there this quantity is unrelated to any 
quantum resolution of the constituents of the metric.}

\section{$1/N$-Corrections Account for Information} 

  Obviously,  the $1/N$-corrections to semi-classical results are much stronger 
  than the naively-expected $e^{-N}$-correction. 
 However, from the fist glance these  enhanced corrections still look very small.  This smallness is an illusion and in reality $1/N$-corrections are precisely what one needs for the correct accounting of information-retrieval in black hole decay.    
  
  The reason is that $1/N$-corrections to planar results are taking place 
  for each act of emission.  Over a black hole half-lifetime this deviation accumulates to order-one effect, which is sufficient to start resolving the information at order-one rate.  As we have shown \cite{Hair},  this reproduces Page's time \cite{Page},
which automatically follows from our picture. 

 It is crucial that $N$ is not a fixed characteristics of the theory 
 (unlike in gauge theories with $N$-colors) but rather a characteristic of a particular black hole.  Moreover, it is a good characteristic only during the 
 time $\sim \sqrt{N} L_P$, during which the black hole depletes and leaks 
 decreasing $N$ by one unit.  This process continues self-similarly    
 \begin{equation}
     N \rightarrow  N -1 \rightarrow  N - 2  .... 
  \label{cascade}
  \end{equation} 
   Each elementary  step of the cascade reveals a distinct feature (information) 
 encoded in a $1/N$-suppressed deviation from the Hawking's  idealized   
 semi-classical result.  To  resolve this feature immediately is extremely 
improbable, but this is not an issue.  Unitarity does not require the information 
to be resolvable immediately.  It only requires that information is resolvable 
on the time-scale of black hole evaporation. 

 This is exactly the case, since probability to recognize the given feature over the 
 black hole half-lifetime, which scales as $ \sim N^{3/2}L_P$, is of order one.  In other words, 
 the increase of $N$ suppresses the probability of decoding a given feature 
 per emission time as $N^{-3/2}$,  but correspondingly  the black hole life-time 
 increases as $N^{3/2}$, so that the product is always of order one.  As a result, 
 for arbitrarily large $N$ the information starts to be recognizable at order-one rate after a half-lifetime of a black hole. 

  To reiterate the picture, let us imagine a situation when Alice 
  is observing  evaporation of a solar mass black hole.  For simplicity, we shall exclude all non-gravitational species from the theory. 
   Then from our point of view,  such a black hole is a BEC of 
 approximately kilometer wavelength gravitons of  occupation 
   number  $N \sim 10^{76}$, with $\sim N$ gapless Bogoliubov modes. 
From the point of view of the quantum information  this black hole 
is a message encoded in a $N \sim 10^{76}$ long sequence of $0$-s and $1$-s, 
\begin{equation}
BH \, = \, (0,0,1,0,1,1, 1, .......), 
\label{message}
\end{equation}
where, the sequence is determined by the state of Bogoliubov modes. 

 After every time interval of approximately  $ \Delta \tau \sim \sqrt{N}L_P \, \sim \, 10^{-5}$ sec
 the message emits a graviton and becomes shorter by one unit. 
  In the semi-classical (planar)  approximation (\ref{planar})  Alice thinks that she sees a thermal evaporation of a black hole with a featureless (exactly thermal) spectrum. 
  However, in reality she sees a depletion and leakage of graviton BEC, with features encoded in sub-leading $1/N$-corrections.  As we know \cite{Nportrait, Hair},  this correction to the black hole  rate goes as 
  $\Gamma_{feature} \, \sim \, N^{-3/2}L_P^{-1}$. Thus, probability for Alice to recognize the feature per emission time is
  $\Delta P =  \Gamma_{feature} \Delta \tau\, \sim 1/N$.  For a solar mass black hole this probability is $10^{-71}$ and is tiny. However, the time-scale available 
  for Alice to resolve the feature is also enormous, and is given by the 
  black hole life-time $\tau = N\Delta \tau \sim N^{3/2} L_P \sim 10^{73}$sec!
  The probability to resolve the feature during this time is
  \begin{equation}
  P \sim  \Delta P N \sim 1\, .
  \label{probability}
  \end{equation}
 Notice, that by then Alice has witnessed $\sim N$ acts of emission and had 
 of order $\tau \sim N^{3/2}L_P$ time for analyzing each of them. Consequently she accumulated order one knowledge about roughly the half of the structure of the message.  This knowledge brings her to the point starting from which she 
 begins to resolve information with order one probability.

 It is important to stress that we are not modifying Hawking's entanglement at each step of the emission process by a factor $\frac{1}{N}ln2$. This would not do the job of reproducing Page's time \cite{Mathur}. What we are instead doing is to use $1/N$ effects ( at the quantum critical point ) to trigger depletion of {\it one} bit of information with probability $1/N$ in {\it each step} of the evaporation process.
 
    This completes our point of nullifying the information paradox.  
   Notice,  that increasing $N$ is not changing the final answer, since although 
   it suppresses the feature per emission time,  it also increases the available 
  time for resolving it  so that the two effects always balance each other.  
  
   This analysis also makes clear the fundamental mistake in the standard semi-classical reasoning. If the features were suppressed by $e^{-N}$ instead of 
   $1/N$, Alice 
   would have never had enough time for resolving these features, and the paradox would follow.  It is now clear that this "paradox" was a result of our misconception about the quantum properties of macroscopic black holes. 
   
   In summary the Bose-Einstein condensate approach to the black hole information paradox lies on the following basic points:
 \begin{itemize}
 \item Black hole emission is due to quantum depletion triggered by quantum noise. This quantum emission is not based on any form of Hawking pair creation in the near horizon geometry. It is a perfectly unitary process with a rate determined by the microscopic dynamics of the condensate.
 
\item This emission rate is modified by $1/N$ effects.
 
\item In particular if we tag a subset of $N_B$ quanta the rate of leakage of any form of information encoded in those quanta ( as could be a  baryon number of the black hole ) goes like $\frac{N_B}{N^{\frac{3}{2}}}$. This in particular means that the black hole can successfully hide some information as its baryon number -- or any other form of message encoded within the tagged quanta -- but only until reaching the half-evaporation point. The {\it observable} prediction of this picture is $1/N$ hair.  In case of baryon number this hair  can have observable effects for astrophysical 
black holes, that are mostly made out of baryons. 

 In this respect  we need to stress the following.  Of course,  one could argue that a general believe that 
unitary quantum gravity should not result in information paradox implicitly assumes that some mechanism should purify Hawking radiation. 
 However,  an issue that has never been addressed previously is  how this  potential purification of Hawking radiation affects the {\it folk dictum} that in any consistent theory of gravity there are no global symmetries. We want to stress that in the Bose-Einstein  portrait approach to the mechanism of information retrieval, gravity is perfectly consistent with global symmetries \cite{Hair}. Obviously, how purification affects the {\it dictum} depends on the strength of the corrections used to purify the emitted quanta.  Our $1/N$-corrections revoke the {\it dictum}.


 \end{itemize}
In short, semi-classicality breaks down whenever quantum noise $1/N$-effects become significant. This is unavoidably the case at the quantum phase transition point. The black hole emits as a normal quantum system, but its identity card is to be at a {\it quantum phase transition point}. 

$~~~~~~~$ 

 Finally, we wish to note on  a  possible avenue of probing the large $N$-picture.   Recently, Veneziano  \cite{venezia} suggested a very interesting stringy computation that reveals $1/N$-hair  in string - string-hole  scattering. 
Viewed as a black hole of occupation number  $N = 1/g_s^2$ ($g_s\equiv$ string coupling),  this result  represents a manifestation of $1/N$-hair suggested by black hole  quantum $N$-portrait.


\section*{Acknowledgements}

 We thank  Gabriele Veneziano for discussions  and for sharing his preliminary results with us. 
 It is a pleasure to thank  Stanley Deser,  Daniel Flassig,  Andrei Gruzinov, 
 Alex Pritzel and Nico Wintergerst for discussions.   
The work of G.D. was supported in part by Humboldt Foundation under Alexander von Humboldt Professorship,  by European Commission  under 
the ERC advanced grant 226371,   by TRR 33 \textquotedblleft The Dark
Universe\textquotedblright\   and  by the NSF grant PHY-0758032. 
The work of C.G. was supported in part by Humboldt Foundation and by Grants: FPA 2009-07908, CPAN (CSD2007-00042) and HEPHACOS P-ESP00346.

\end{document}